\documentclass[useAMS,usenatbib]{mn2e}

\usepackage{graphicx}
\usepackage{times}
\usepackage{aas_macros}
\title[CR escape from SNRs]{Escaping the accelerator; how, when and in what numbers do cosmic rays get out of supernova remnants?}
\author[L. O'C. Drury]{L. O'C. Drury\\
School of Cosmic Physics, Dublin Institute for Advanced Studies, 31 Fitzwilliam Place, Dublin 2, Ireland}

\begin{document}
\maketitle

\begin{abstract}
The escape of charged particles accelerated by diffusive shock acceleration from supernova remnants is shown to be a more complex process than normally appreciated.  Using a box model it is shown that the high-energy end of the spectrum can exhibit spectral breaks even with no formal escape as a result of geometrical dilution and changing time-scales.  It is pointed out that the bulk of the cosmic ray particles at lower energies must be produced and released in the late stages of the remnant's evolution whereas the high energy particles are produced early on; this may explain recent observations of slight compositional variations with energy.  Escape resulting from ion-neutral friction in dense and partially ionized media is discussed briefly and some comments made on the use of so-called ``free escape boundary conditions''.
Finally estimates are made of the total production spectrum integrated over the life of the remnant.
\end{abstract}


\section{Introduction}

The idea that Supernova Remnants (SNRs) produce the bulk of the Galactic cosmic rays is now widely accepted.    A key element in this general acceptance has been the development of a quite sophisticated theory of particle acceleration in shock waves, usually called diffusive shock acceleration (DSA), which provides a compelling explanation of how a small number (of order $10^{-4}$ for conventional SNR parameters) of particles flowing into the shock can be accelerated to ultra-relativistic energies and become cosmic ray particles.   Less attention has been paid to the question of how these particles then escape from the acceleration site and propagate into the interstellar medium although this is clearly an important question.  However increasing attention is now being paid to this issue, in part because of theoretical developments relating to magnetic field amplification which require it to be taken much more seriously, and in part because of interest in the idea that escaping particles illuminating nearby molecular clouds might be sources of high energy gamma-rays and neutrinos \citep{2010PASJ...62..769C,2007ApJ...665L.131G,2007Ap&SS.309..365G}

\section{Theoretical preliminaries}

In the standard simple theory of DSA particles are accelerated while within a diffusion length of the shock, there is no escape of particles upstream whatsoever, and the only escape downstream is by advection with the bulk flow 
\citep{1977DoSSR.234Q1306K,1977ICRC...11..132A,1978MNRAS.182..147B,1978ApJ...221L..29B}
As long as the shock can be treated as an essentially planar structure propagating at fixed speed into an upstream medium capable of scattering particles this picture is certainly correct.  Asymptotically a plane surface moving at uniform speed, where the displacement is proportional to the time, will always overtake a randomly walking particle where the displacement grows only as the square root of the time.  Thus at this level all particles are accumulated in the downstream region and there is no escape into the undisturbed far-upstream medium.  

The problem of course is that a lot of simplifying assumptions have been made in this picture.  A real SNR shock is spherical and not planar; decelerating and not moving at constant speed; and the upstream scattering will certainly depend on the effective magnetic field strength as well as the wave spectrum and may even be totally suppressed in partially ionized regions of the ISM. The picture is further complicated when the nonlinear reaction of the accelerated particles on the shock structure, and the resulting secular evolution of the shock structure, is included.  Of course, and relevant to this paper, if the scattering is totally suppressed particles can escape upstream even from a uniformly moving and infinitely extended planar shock, but it is important to note that this is a singular limit.  Letting the shock radius go to infinity at finite scattering and letting the scattering go to zero at finite radius are two mathematical limiting processes that do not commute, so the zero scattering and infinite shock problem is formally ill-posed.   For the problem of escape from supernova remnants what we actually want to study is the case of a large but finite shock radius and an upstream medium (the ISM in our Galaxy) which has a non-zero even if small level of scattering (as required by cosmic ray propagation models).

Naively one may expect that, because $0.5$ is larger than $0.4$, a spherical blast wave expanding according to the Sedov scaling, $R\propto t^{0.4}$, between radius $R$ and time $t$ can be outrun by a randomly walking particle with displacement $D$ growing as $D\propto t^{0.5}$ so that escape has to be considered a real possibility.  In fact a more sophisticated argument is required because the particle in the course of its random walk may return to the shock even though on average it moves further away.  It is in fact quite easy to show (see appendix) that a uniformly diffusing particle released at radius $R_1$ from the origin will enter a sphere centred on the origin of radius $R_0$ with probability $R_0/R_1$ and escape to infinity without ever encountering the sphere with probability $1-R_0/R_1$.  Thus a particle located even as far upstream as ten shock radii will have a slightly larger than 10\% chance of returning to the shock (slightly larger because the shock is expanding; it would be precisely 10\% for a stationary spherical surface).  This is of course a mathematical result which among other things depends on uniform and isotropic spatial diffusion in three spatial dimensions and as such must be treated with appropriate caution, but it does show that escape is not quite as straightforward as one might think.  Further while escape is possible in three dimensions it is formally impossible in one dimension; a particle randomly walking on a line visits every point infinitely often even though the recurrence intervals grow ever longer.  Thus if the cosmic rays were effectively tied to a single magnetic field line they could never escape (at least formally).  In reality of course there is some element of cross-field diffusion and the three dimensional result is probably more relevant.

Even it there is a non-zero probability of return to the shock, for acceleration to continue this probability has to be very close to one (more precisely, it must be less than one by an amount of order $U/c$ where $U$ is the shock speed and $c$ is the velocity of light).  Thus there can be a large intermediate range where particles are no longer being accelerated, but have not really escaped from the shock either (if one defines this as having negligible probability of coming back to the shock front).  This can be shown more precisely using the so-called box model as was done in 
\cite{2003astro.ph..9820O}.  For convenience we summarise the key points here while slightly generalising the treatment.

\subsection{An example of acceleration spectral breaks without escape in the ``box'' model}
The simplest approximate, but quite physically realistic, treatment of shock
acceleration is the so-called ``box'' model \citep{1999A&A...347..370D}. In this the accelerated
particles are assumed to be more or less uniformly distributed
throughout a region extending one diffusion length each side of the
shock, and to be accelerated upwards in momentum space at the shock
itself with an acceleration flux
\begin{equation}
\Phi(p) = {4 \pi\over3} p^3 f(p) \left(U_1 - U_2\right),
\end{equation} 
per unit surface area
where $U_{1}$ and $U_{2}$ are the upstream and downstream velocity and $f(p)$ 
is the phase space density of the accelerated particles (assumed to have an \
almost isotropic distribution).
If the diffusion
length upstream is $L_1$, and that downstream is $L_2$, then
\begin{equation}
L_1 \approx {\kappa_1(p)\over U_1},\qquad L_2 \approx {\kappa_2(p)\over U_2},
\end{equation}
where $\kappa_1$ and $\kappa_2$ are the upstream and downstream
diffusion coefficients. To a first approximation we assume that both
$L_1$ and $L_2$ are small relative to the radius of the shock and that
we can neglect effects of spherical geometry (in fact it is not too
difficult to develop a spherical box model, but it unnecessarily
complicates the argument) so that the box volume is simply
$A(L_1+L_2)$ where $A$ is the surface area of the shock.
The basic ``box'' model equation is then simply a conservation
equation for the particles in the box; the rate at which the number in
the box changes is given by the divergence of the acceleration flux in
momentum space plus gains from injection and advection and minus
advective losses to the downstream region.
\begin{eqnarray}
{\partial\over\partial t} \left[ A (L_1 +  L_2) 4\pi p^2 f(p) \right]
&+& A{ \partial \Phi\over\partial p} = A Q(p) \nonumber\\
&+& A F_1(p) - A F_2(p),
\end{eqnarray}
where $Q(p)$ is a source function representing injection at the shock
(only important at very low energies), $F_1$ is a flux function
representing advection of pre-existing particles into the system from
upstream (normally neglected) and $F_2$ is the flux of particles
advected out of the system and carried away downstream.  The only
complication we have to consider is that the box is time-dependent,
with flow speeds, shock area and diffusion lengths all changing.

The escaping flux is determined simply by the advection across the 
downstream edge of the box, that is
\begin{equation}
F_2(p) = 4 \pi p^2 f(p) \left(U_2 - {\partial L_2\over\partial
  t}\right),
\end{equation}
where we have to explicitly allow for the time varying size of the
downstream region.
Substituting this expression for $F_2(p)$ and neglecting the advection of prexisting particles 
(the $F_1(p)$ term) the box equation
simplifies to:
\begin{eqnarray}
{1\over A} {\partial A\over\partial t} \left( L_1+L_2 \right) f
&+& {\partial L_1\over\partial t} f
+\left(L_1+L_2\right){\partial f\over\partial t} 
+ U_1 f \nonumber\\
&+&\left(U_1 - U_2\right) {p\over 3} {\partial f\over\partial p} =
{Q\over 4\pi p^2}.
\end{eqnarray}

Partial differential equations of this form always reduce, by the
method of characteristics, to the integration of two ordinary
equations, one for the characteristic curve in the $(p,t)$ plane
\begin{equation}
\label{char}
{{\rm d}\,p\over {\rm d}\,t} = {U_1 - U_2 \over L_1 + L_2} {p\over3},
\end{equation}
and one for the variation of $f$ along this curve
\begin{eqnarray}
\left(L_1 + L_2\right) {{\rm d}\,f\over {\rm d}\,t} 
&+& f\left[\left(L_1+L_2\right){1\over A}{\partial A\over\partial t}
+{\partial L_1\over\partial t} + U_1\right] \nonumber\\
&=& {Q\over 4\pi p^2}.
\end{eqnarray}                                                                
$Q=0$ everywhere except at the injection momentum and we can write the
above equation as
\begin{equation}
\label{PDE}
{{\rm d}\, \ln f\over {\rm d}\, t}
= - {1\over A}{\partial A\over\partial t}
  - {1\over L_1+L_2} {\partial L_1\over \partial t}
  - {U_1\over L_1 + L_2}.
\end{equation}
But the
shock area $A$ is a function only of time so that 
\begin{equation}
{\partial A\over\partial t} = {{\rm d}\, A\over {\rm d}\, t},
\end{equation}
and, although the upstream diffusion length does depend on both time
and momentum, if we assume Bohm scaling for the two lengths so that
\begin{equation}
L \propto {\kappa\over U}  \propto {p v\over U B},
\end{equation}
(where $v$ is the particle velocity) we can write
\begin{equation}
{1\over L_1+L_2} {\partial L_1\over \partial t}
= - \vartheta {{\rm d}\,\ln(U_1 B_1)\over {\rm d}\, t},
\end{equation}
where
\begin{equation}
\vartheta = {L_1\over L_1 + L_2},
\end{equation}
(obviously $0<\vartheta<1$).
Finally, noting that
\begin{equation}
{U_1\over L_1 + L_2} = {3 U_1\over U_1 - U_2} {{\rm d}\,\ln p\over 
{\rm d}\, t},
\end{equation}
and a fixed compression
ratio, we can simplify equation (\ref{PDE}) to 
\begin{equation}
{{\rm d}\, \ln f\over {\rm d}\, t} = - {{\rm d}\, \ln A\over {\rm d}\, t}
+ \vartheta {{\rm d}\,\ln(U_1 B_1)\over {\rm d}\, t}
-  {3 U_1\over U_1 - U_2} {{\rm d}\,\ln p\over {\rm d}\, t},
\end{equation}
which integrates trivially to relate the value of $f$ at the end of
one of the characteristic curves, say at the point $(p_1, t_1)$, to the
value at the start, say at $(t_0, p_0)$, as follows;
\begin{equation}
\label{int1}
{f(t_1, p_1)\over f(t_0, p_0)} =
\left(A(t_1)\over A(t_0)\right)^{-1} 
\left(U_1(t_1) B_1(t_1)\over U_1(t_0) B_1(t_0)\right)^\vartheta
\left(p_1\over p_0\right)^{-s}
\end{equation}
where
\begin{equation}
s = {3U_1\over U_1-U_2}
\end{equation}
is the standard exponent of the steady-state power-law spectrum associated with
shock acceleration.

This rather beautiful result shows how the standard test particle
power-law is modified by a combination of effects as the box volume
changes. As one would expect the amplitude varies inversely as the
shock area and also decreases if the upstream diffusion length (at
fixed energy) increases, but with an exponent between zero and one
determined by the ratio of the upstream diffusion length to the total
width of the diffusion region. It is very interesting that the result
is not simply a variation inversely as the box volume, which one would
naively expect from geometrical dilution. This reflects the
fundamental asymmetry between the upstream and downstream regions,
that upstream is empty outside the diffusion region whereas the entire
downstream region is filled with accelerated particles.

If we assume pure Bohm scaling the other differential equation is also
integrable so that the problem is reduced entirely to quadratures (of
course only within the various approximations we are making; but still
a remarkable result).  Bohm scaling implies that the mean free path is
of order and proportional to the particle gyroradius, so that if the
particle charge is $e$
\begin{equation}
L_1 + L_2 \approx \alpha {p v \over 3 e B_1 U_1},
\end{equation} 
where $\alpha$ is a dimensionless parameter (of order ten).
Substituting in the equation of the characteristic (equation (\ref{char}))we get
\begin{equation}
v {{\rm d}\, p\over {\rm d}\, t} = {1\over\alpha} \left(U_1-U_2\right) U_1 e B_1,
\end{equation}
and noting the relativistic identity between kinetic energy $T$,
momentum $p$ and velocity $v$,
\begin{equation}
v = {{\rm d}\, T\over {\rm d}\, p},
\end{equation}
we can integrate this as
\begin{equation}
\label{kinen}
T_1 - T_0 = {e\over\alpha} \int_{t_0}^{t_1} \left(U_1-U_2\right)U_1 B_1 \,dt.
\end{equation}
For relativistic particles the kinetic energy and the momentum are
essentially interchangeable with $T= c \sqrt{p^2 + m^2 c^2} - m c^2
\approx c p$.

Further progress can be made by assuming that the shock velocity follows the Sedov self-similar law for a strong spherical explosion in a cold gas
where the shock radius expands as $R\propto t^{2/5}$ and the shock velocity
decreases as $U\propto t^{-3/5}$.  We further suppose that the effective magnetic field scales as a power of the shock velocity,
\begin{equation}
B_{\rm eff} \propto U^\mu\propto t^{-3\mu/5}.
\end{equation}
The parameter $\mu$ characterises the degree of field amplification in the shock.  No amplification corresponds to $\mu=0$, energy equipartition would give $\mu=1$, and Bell's instability 
\citep{2004MNRAS.353..550B} gives $\mu=1.5$.  With these power-law scalings the integration is trivial and yields the acceleration paths,

\begin{eqnarray}
T_1 - T_0 
&\propto& \int_{t_0}^{t_1} (U_1-U_2)U_1B dt\nonumber\\
&\propto& \int_{t_0}^{t_1} t^{-(6+3\mu)/5} dt\nonumber\\
&\propto& t_0^{-(1+3\mu)/5} - t_1^{-(1+3\mu)/5}.
\end{eqnarray}

These curves all rise extremely
steeply, representing an initial phase of rapid acceleration, turn
over, and then become asymptotically flat.  Physically it is clear
that, as the shock slows and the field drops, the high energy
particles cease to be significantly accelerated and simply diffuse
further and further in front of the shock. 

A very important aspect of the curves is that they
uniquely relate final energies (or equivalently momenta) to starting
times. Asymptotically and at high energies the relation is a simple power-law; for $T_1 \gg
T_0$ and $t_0 \ll t_1$ we have simply
\begin{equation}
p_1 \propto T_1 \propto t_0^{-(1+3\mu)/5}, \qquad t_0 \propto p_1^{-5/(1+3\mu)}.
\end{equation}
Using this we can now translate the dilution factors from
equation (\ref{int1}) to additional power-law terms in the final momentum. 
Explicitly, a given final momentum maps to a starting radius using a 
Sedov expansion-law:
\begin{equation}
R(t_0) \propto t_0^{2/5} \propto p_1^{-2/(1+3\mu)},
\end{equation}
and thus the first term on the RHS of equation (\ref{int1}) translates to a:
\begin{equation}
\label{RHS1}
\left(A(t_1)\over A(t_0)\right)^{-1}  \propto R(t_0)^2 \propto p_1^{-4/(1+3\mu)}
\end{equation}
(note that we are thinking of the spectrum as a function of final momentum $p_1$ at some fixed final time $t_1$, so that $A(t_1)$ is constant).
The final momentum can also be mapped to a starting velocity, using again
a Sedov expansion-law:
\begin{equation}
U(t_0) \propto t_0^{-3/5} \propto p_1^{3/(1+3\mu)}.
\end{equation}
Thus the second term on the
RHS of equation (\ref{int1}) scales as:
\begin{equation}
\label{RHS2}
\left(U_1(t_1) B_1(t_1)\over U_1(t_0) B_1(t_0)\right)^\vartheta\propto
\left(U_1(t_1)\over U_1(t_0)\right)^{(1+\mu)\vartheta}
\propto p_1^{-{3(1+\mu)\vartheta\over 1+3\mu}}
\end{equation}

Thus for the high-energy part of the spectrum we deduce that
\begin{equation}
f(p_1) \propto f(t_0, p_0) \left(p_1\over p_0\right)^{-3U_1\over U_1 - U_2} p_1^{-{4+3\vartheta(1+\mu)\over 1 + 3\mu}}
\end{equation}

Finally  we need to determine the initial amplitude of $f$ from an
injection model.   There are two main approaches to the injection rate.  The first,
which is perhaps more consistent with the test particle approach, is
to simply parametrise it by assuming that some fraction of the
incoming thermal particles are ``injected'' as non-thermal particles
at some suitably chosen ``injection momentum'' which 
separates the thermal particle population from the non-thermal.  
In other words one writes
\begin{equation}
Q(p,t) = \eta(t) n_1 U_1 \delta(p-p_{\rm inj}(t)),
\end{equation} 
where $n_1$ is the upstream thermal particle number density, $\eta\ll 1$ is
the injection fraction, $p_{\rm inj}$ is the injection momentum and as
usual $\delta$ is Dirac's delta distribution.  It
should be clear that this is a parametrisation rather than a true
injection model, however it, or equivalent parametrisations, have been
very widely used, typically with $\eta$ taken to be a constant of
order $10^{-5}$ to $10^{-4}$ for protons and $p_{\rm inj}\approx 10
m_p U_1$ where $m_p$ is the proton mass.  However there is no real
justification for this apart from the fact that it seems to yield
reasonable results in many cases.

With the above parametrisation the distribution function just above
the injection energy can be simply determined by equating the
acceleration flux to the injection flux,
\begin{equation}
{4 \pi p_{\rm inj}^3\over 3} \left(U_1-U_2\right) f(p_{\rm inj}) 
= \eta n_1 U_1,
\end{equation}
giving
\begin{equation}
f(p_{\rm inj}) = {3\over 4 \pi p_{\rm inj}^3} {U_1\over U_1 - U_2} n_1
\eta.
\end{equation}

On this injection model, noting that the injection momentum scales as the initial shock velocity, thus $p_0\propto U \propto p_1^{3/(1+3\mu)}$, it is easy to see that the power-law exponent of the spectrum at high energies is modified to
\begin{equation}
{\partial \ln f\over\partial \ln p} =
- {3U_1\over U_1-U_2} 
+ {3\over 1+3\mu} {3U_2\over U_1-U_2} - {4 + 3\vartheta(1+\mu)\over 1 + 3\mu}
\end{equation}
The first correction is essentially a hardening of the spectrum reflecting increased injection at early times while the second is a softening resulting from the combined dilution terms.

The second approach adopts the idea, which can be traced back to the
early work of \cite{1979ApJ...229..419E}, that the injection process is inherently
extremely efficient but that various feedback processes
operate to reduce it to the point where the accelerated particles 
carry a significant part of the energy dissipated in the
shock.  For
a standard spectrum close to $p^{-4}$ the energy is almost uniformly
distributed per logarithmic interval over the relativistic part of the
spectrum.  This suggests taking a reference momentum in the
mildly relativistic region, $p_0\approx m c$, and determining $f$ by
a relation of the form
\begin{equation}
{4\pi p_0^3\over 3} f(p_0) m c^2 \approx {1\over\lambda} \rho U_1\left(U_1-U_2\right)
\end{equation}
where $\lambda$ is a number which depends logarithmically on the upper
cut-off and which for supernova remnants is probably somewhere between
$10$ and $100$.   

On this injection model the reference momentum is fixed and the amplitude of the spectrum scales with the ram pressure of the shock,
\begin{equation}
f(p_0) \propto \rho U_1(U_1-U_2) \propto p_1^{6/(1+3\mu)}.
\end{equation}
It follows that the high-energy asymptotic slope is modified to
\begin{equation}
{\partial \ln f\over\partial \ln p} =
- {3U_1\over U_1-U_2}  + {2 - 3\vartheta(1+\mu)\over 1 + 3\mu}
\end{equation}
Again because of the high injection rate at early times the high-energy spectrum can even flatten, at least in theory.   However it is important to note that this calculation is not strictly consistent in that we are assuming that the injection is sufficiently strong for non-linear effects to become important, but are ignoring the reaction effects on the shock structure in the acceleration calculation.    The assumption that the diffusion length scales are small relative to the shock radius also breaks down and this will increase the importance of the dilution effects which are underestimated here.
{\em Thus both calculation should be viewed as toy models illustrating some of the effects that can occur rather than as definitive predictions.}

A further caveat is that both injection models are
models for proton injection, the protons being the dynamically
dominant species. Unfortunately very little is known about the factors
controlling the injection of electrons and other minor species despite
their importance for diagnostic tests. It has also been suggested that
the injection may be nonuniform over the shock surface with a strong
dependence on the angle between the mean background field and the
shock normal \citep{2003A&A...409..563V}.

These results refer of course only to the asymptotic behaviour of the high
energy part of the spectrum. As $p_1$ is decreased there comes a point
where $t_0$ is no longer small relative to $t_1$. At this point all
values of the final momentum map down to a small approximately
constant region and the spectrum becomes simply the standard 
spectrum with exponent $-3U_1/(U_1-U_2)$.  This break occurs at the point to which efficient acceleration is possible at
that stage in the remnant evolution, and decreases as the the remnant ages. 

This calculation illustrates an important point.  The break in the spectrum associated with the acceleration time scale becoming of order the dynamical time scale can be at a significantly lower energy than that at which particles are escaping from the shock and it is quite possible for a population of energetic particles to remain in the neighbourhood of the shock, and even return to it occasionally, without significant acceleration.    This population will be geometrically diluted by the expansion of the shock and the increase in the upstream diffusion lengh (the region of space they fill continues to expand faster than they can be supplied by the on-going acceleration) and this process can rather naturally produce spectral breaks at the point where the acceleration is no longer in equilibrium with the expansion.  The break in the spectrum reflects not just the geometrical dilution but also, and potentially of great interest, the time-history of injection at the shock.

This effect also explains an apparent paradox in the simple theory.  If the maximum particle momentum is calculated as
\begin{equation}
p_{\rm max} = \int_0^t {p\over t_{\rm acc}(t')} dt'
\label{Max1}
\end{equation}
where $t_{\rm acc}$ is the acceleration time scale it is clearly a monotonic increasing function of time.  However if one estimates the maximum momentum by arguing that the diffusion length has to be less than the shock radius, or equivalently that the local acceleration time has to be less that the dynamical time, this gives the condition
\begin{equation}
t_{\rm acc} \approx 0.1 {\kappa(p_{\rm max})\over\dot R^2}< R/\dot R
\label{Max2}
\end{equation}
where $R$ is the shock radius and $\dot R$ its expansion speed.  Thus on this condition the maximum momentum is implicitly determined by
$\kappa(p_{\rm max}) \approx 0.1 R\dot R$ 
For a Sedov blast wave the product $R\dot R\propto t^{-0.2}$ is a decreasing function of time (even if slowly) and as $\kappa$ is an increasing function of $p$ one expects that $p_{\rm max}$ should actually decrease with time and not increase.   The solution of course is that the spectrum has a break at the point given by equation (\ref{Max2}) (where the acceleration is no longer able to be in equilibrium with the expansion), but continues to the higher energy given by the first condition of equation (\ref{Max1}).

To some extent it thus comes down to semantics.  One can have a perfectly consistent picture of active SNRs being surrounded by a halo of particles which were accelerated at earlier times to high energies, but which are no longer undergoing significant acceleration and are simply diffusing out into the Galaxy.  But of course if this halo region were to become a large part of the Galaxy, and the probability of further interaction with the shock vanishingly small, then clearly it would make more sense to regard these simply as escaped particles.  And of course once the shock becomes weak and the SNR effectively dies then all the particles, including those trapped in the interior, must eventually escape.  Although very uncertain it is interesting to make some order of magnitude estimates. If the diffusion into the Galaxy occurs as indicated by studies of cosmic ray propagation, with a diffusion coefficient of order $10^{28}\,\rm cm^2s^{-1}$ for particles around a GeV and an energy dependence something like $E^{0.5}$, then PeV particles from a thousand year old SNR would diffuse a distance of order $\sqrt{3\times 10^{41}}\,\rm cm\approx 180\, pc$.  Thus particles produced in a hypothetical early pevatron phase of a historical SNR will not have escaped into the general Galaxy but will fill a roughly spherical region around the SNR of radius a few hundred parsecs.  This is of course the motivation for looking for high-energy emission from molecular cloud targets near SNRs.

\section{Trapping inside the shock of low-energy particles}

In fact for the bulk of the cosmic rays, and in particular for those which dominate the energy density and for which the composition is well determined, trapping until the SNR dies is undoubtedly the better picture.  At these relatively low energies the planar shock approximation is good, the particles are accelerated at the shock and then swept downstream to accumulate in the interior of the remnant.  Here of course they undergo adiabatic energy losses as the remnant expands, but the energy lost in this way goes to driving the shock and is thus recycled into the acceleration of new particles.

The acceleration depends crucially on having strong scattering in the neighbourhood (and in particular upstream) of the shock which produces a diffusion barrier at the shock preventing the escape of these low-energy particles.  There are a wide variety of resonant and non-resonant instabilities which can produce the necessary magnetic turbulence and it is generally believed that the local diffusion coefficient at the shock can be driven down to values corresponding to Bohm scaling, that is a scattering mean free path of order the gyroradius.  The sharp synchrotron rims observed in young remnants, originally in the radio \citep{1994A&A...281..220A} and more recently in X-rays \citep{2004A&A...416..595Y}
provide rather convincing direct observational evidence for such diffusion barriers associated with strong shocks.  Were the diffusion typical of that inferred in the general ISM the rims would be much more extended and indeed as pointed out long ago shock acceleration would not be a viable mechanism for making the Galactic cosmic rays, e.g. \cite{1981ICRC....2..336G}.  Eventually however the shock will become too weak to sustain the required strong scattering, the diffusion barrier will collapse and at this point the particles in the interior will escape into the ambient medium as the shock (and by implication the SNR) dies.

\subsection{Energy-dependent composition changes}

This has the important consequence that at lower energies the cosmic ray composition should be dominated by particles accelerated just before the remnant dies (thereby suppressing any freshly synthesised component coming from the SN or its progenitor wind).  This was discussed in \cite{1995NuPhS..39..165D}.  The key point is that, as discussed above, the low energy particles trapped in the interior are subject to adiabatic losses on the dynamical time scale of the remnant, and that this energy is then recycled into producing freshly accelerated low-energy particles. 
Thus the dominant low-energy particle population is that accelerated within the most recent dynamical time-scale of the remnant.  This energy argument is the more fundamental one, but in fact even if one simply looks at the amount of matter swept up, far more matter is swept up at late times when the remnant is large than the amount (of order the ejecta mass) swept up prior to the onset of the Sedov phase.

It is interesting to note that the picture is very different for the highest energy particles which have to be accelerated at the time when the shock is fastest (probably in the period leading up to the transition from free expansion to the Sedov phase) and it is tempting to speculate that this should produce slight compositional variations with energy.  
Indeed there appear to be such variations in recent experimental data, and we may well be seeing evidence of this effect.   To put it succinctly, the composition at high energies should reflect acceleration early on prior to the mass sweep-up time when the ejecta interact with the same mass of swept-up material; the composition at low energies should reflect acceleration at the energy sweep-up time when the total thermal energy of the swept-up ambient material becomes comparable to the explosion energy (and the shock is weakening).  In particular the apparently softer proton spectra compared to helium can be very easily explained by the older and weaker shocks propagating in a medium where more of the helium is neutral than in the strongly ionized environment of a young remnant.

\section{An empirical estimate of magnetic field amplification}

As indicated in the introduction one reason that it is topical to discuss these issues is because of the recent interest in magnetic field amplification in SNR shocks.  Without field amplification the difference between the two maximum momenta given by equations \ref{Max1} and \ref{Max2} above is quite small (at least for normal SNR parameters) and the discussion could be regarded as rather academic.  The situation is radically different if the effective magnetic field strength is a strongly increasing function of shock speed.  In this case the local equilibrium energy becomes a strongly falling function of time and the gap between the two maximum energies can be many decades.  Similar discussions can be found in \citet{2009MNRAS.396.1629G,2008ApJ...678..939Z}.
In fact if one wishes to explain the particles between the ``knee'' in the cosmic ray spectrum at about $3\,\rm PeV$ and the ``ankle'' at $3\,\rm EeV$ in this way, then at least two decades have to be bridged in this way (chemistry can account for about one decade with the Iron spectrum extending to a higher total energy per particle than the proton spectrum).  Thus observationally we require that the product $BR\dot R$ be of order $10^{17}\,\rm V$ in the strong young shocks accelerating the highest energy particles and fall to a value more like $10^{15}\,\rm V$ just before the shock dies.  

This two decades decrease must be almost entirely in the magnetic field strength, which must therefore be quite strongly dependent on the shock speed.  Naively one may argue that the strongest shocks have speeds of order $10^4 \,\rm km\,s^{-1}$  whereas the weakest are probably of order $10^2\,\rm km\,s^{1}$ so that a roughly linear dependence fits well - but of course this is a very uncertain and crude estimate.

\section{Suppresssion of scattering in dense media}

A further complication is that the strong scattering required to maintain the diffusion barrier at the shock may fail in dense and partially ionized media because ion-neutral friction suppresses the instabilities needed to disturb the magnetic field on the relevant scales; a good recent discussion is that by  \cite{2005A&A...429..755P}.   This was actually pointed out first in the seminal paper by \cite{1978MNRAS.182..147B}.  Bell, and following him \cite{1993ARA&A..31..373D} assumed that at a finite distance in front of the shock the damping would totally suppress the wave scattering and that particles would then freely stream away at a fraction of the speed of light.  Unfortunately both seriously overestimated the escaping flux by assuming that the intensity at the point where escape sets in would be the same as in the diffusion equation solution with scattering throughout the upstream region.  As pointed out in \cite{1996A&A...309.1002O} one should actually solve the diffusion equation with a modified boundary condition of $f=0$ at the point where free streaming sets in which reduces the flux by a factor of order $U/c$.  A similar transition between a diffusion regime and a free streaming regime occurs in the theory of stellar atmospheres where this effect is well known.  

More recently  \cite{2005ApJ...624L..37M} have advocated an interesting variant of this idea where instead of suppressing all the waves they point out that the damping will selectively suppress a range of wave-numbers which in turn produces an escape cone of finite opening angle in pitch-angle space.   The authors consider the case of particles from a shock precursor beginning to enter a dense molecular cloud and argue that in the cloud there is a critical momentum $p_1$ (their notation, and not to be confused with $p_1$ in the rest of this paper) such that particles with parallel momentum $\left| p_\parallel \right | > p_1$  are not scattered and escape at the speed of light.  They assume that the effect of this is that only particles with $\left| p_\parallel \right | < p_1$
remain in the cloud to produce gamma-rays and that this automatically leads to a steepening of the spectrum by precisely unity because the portion of phase space left is just $p_1/p$.
However if a significant part of the distribution function is removed, one cannot assume that the rest of the distribution function remains unaffected.  In reality particles will rapidly pitch-angle diffuse into the loss-cone from parts of the distribution where the scattering is still strong and pull the distribution down towards zero.  Malkov (personal communication) has clarified that they regard the effect as a rapid and transient one, but it is far from clear that the time-scale for penetration of the precursor into the cloud is short relative to the pitch-angle relaxation time of the distribution. Nor is it clear what effect spatial variation of $p_1$ and mirroring associated with spatial variation of the magnetic field strength have on the proposed mechanism.  

\section{A comment on free escape boundaries}

Although as argued above free-escape boundaries are expected to occur naturally in partially ionized media,  they were originally introduced in discussions of shock acceleration for a totally different reason, namely as an artificial regularising procedure in the non-linear theory.   The Monte-Carlo simulations of non-linear shocks pioneered by Ellison and his coworkers, at least in the  versions developed to date, require the modified shock structure to be planar and stationary.  So also do the semi-analytic theories developed by \cite{1979ApJ...229..419E}, \cite{1997ApJ...485..638M}
 and Blasi which depend crucially on the assumption of stationarity and planarity.  As is well-known a steady modified planar shock produces such hard high-energy spectra that the accelerated particle pressure will diverge unless some cut-off is imposed (even a test-article $f(p)\propto p^{-4}$ spectrum will diverge logarithmically at high energies, and the non-linear effects make this much worse).
In reality the location and shape of the cut-off is determined by factors such as the finite age and size of the shock, but in the calculations all of these are lumped into one artificially imposed cut-off  which is then adjusted to fit the system being studied.  

One way of imposing a cut-off is simply to assume that particles vanish from the system ("escape") when they reach some maximum cut-off energy \citep{1984ApJ...286..691E}.  This has the advantage of simplicity, but looks artificial and gives abrupt step-function cut-offs.  An alternative which is widely used is to impose a so-called "free-escape" boundary at a finite distance upstream of the shock.  The main advantage of this is that it gives a more natural-looking smooth cut-off and in addition, as argued by 
\cite{2009ApJ...694..951R}, it is more physical in that the local diffusion barrier at the shock falls off at a finite distance upstream if there is strong local field amplification.  This paper also gives a useful discussion and comparison of the two approaches.  For a good recent discussion in the context of nonlinear shocks see \citet{2010APh....33..307C}.  A final point worth making is that steady models are by their nature incapable of including the geometrical dilution effects which we have seen to be important in determining the high-energy end of the spectrum.   The dilution effects have to be approximated as escape, which leads to an over-estimate of the actual escape.

While there is nothing wrong with the use of such artificial cut-offs, and indeed they are essential for the success of both the Monte-Carlo and semi-analytic models, their wide-spread use has perhaps contributed to impression that there is a strict dichotomy between particles that are being accelerated and particles that have escaped.  The reality for high-energy particles, as argued above, is that there is no such sharp transition.  Rather there is a gradual slowing of the acceleration and an increasing amount of time spent diffusing ever further upstream, at least as long as there is some finite level of scattering throughout the upstream region.  It does of course depend on the level of upstream scattering and the scales that one is interested in.  If the shock is small and the upstream scattering very weak, as is the case for particle acceleration at the Earth's bow shock for example, a description in terms of escape is entirely appropriate.  If the scattering is almost entirely due to excitation of instabilities by the accelerated particles, and the pre-existing level of turbulence is very low, it is clear that a certain flux of escaping particles is needed to bootstrap the whole process (as observed in PIC simulations) and in this situation nonlinear effects could lead to a rather sharp transition in energy between trapped particles and an escaping population.  Similar considerations apply to all numerical studies where the limited size of the computational box inevitably leads to the need to allow for escape.  But as estimated above, for a SNR expanding into the ISM, the PeV particles even after a thousand years have probably only diffused a few hundred parsecs in the tangled ISM field whereas the SNR can have a radius of several parsecs to tens of parsecs; here it seems more appropriate to think of the accelerated particles as gradually falling out of equilibrium with the shock acceleration and filling an extended halo than as abruptly escaping.

\section{The total source spectrum}

From the point of view of cosmic ray propagation theory what is actually required is an estimate of the source function for cosmic rays, in other words what is the total contribution per SNR to the galactic cosmic ray population?  Leaving aside the complication of interacting remnants in superbubbles (although this must certainly occur, and indeed is indicated by the compositional data) we can try to answer this using the ideas developed above.  A very similar discussion can be found in \cite{Caprioli2010160}; see also \cite{2010A&A...513A..17O}.

Let us consider first the high-energy part of the spectrum (which we take to be the part between the `knee' at about $3\times 10^{15}\,\rm eV$ and the `ankle' around $3\times 10^{18}\,\rm eV$).  The former, on the views presented here, is accelerated early on while the shock is fast and has a strongly amplified field.  As the shock slows and the effective field decreases, the critical momentum, $p_*$ at which the acceleration time-scale is comparable to the dynamical time-scale decreases and particles above this energy stop being accelerated and start to diffuse away from the remnant.
For these particles the production rate is composed of two parts, the acceleration flux at the shock itself upwards through $p_*$ and the numbers of particles already accelerated which are left above $p_*$ as the cut-off momentum falls.  These are respectively
\begin{eqnarray}
Q_1 &=& {4\pi p_*^3\over 3} f(p_*) (U_1-U_2) 4\pi R^2\\
Q_2 &\approx & - 4\pi p_*^2 f(p_*) \dot p_* {4\pi R^3\over 3}\\
\end{eqnarray}
where the second is an approximate estimate only (it assumes that the downstream particles uniformly fill the interior which is certainly not the case, but it also neglects the contribution from the upstream population which should at least partially compensate for this).  The ratio of the two terms is just
\begin{equation}
{Q_1\over Q_2} = \left(-p_*\over \dot p_*\right) \left(U_1- U_2\over R\right)
\end{equation}
which, for the self-similar power-law evolution we are assuming with $p_*\propto t^{-(1+3\mu)/5}$, is just
\begin{equation}
{Q_1\over Q_2} = {2\over 1+3\mu} {U_1-U_2\over U_1}.
\end{equation}
Thus, as is clear on physical grounds, the first term dominates if there is no field amplification but the second term rapidly becomes more important as the field amplification increases.  In general the two terms are of roughly equal importance.

Particles at the critical momentum are produced for a period inversely proportional to the rate at which $p_*$ is decreasing, $dt = - dp_*/\dot p_*$, and thus the spectrum released into the Galaxy is
\begin{equation}
S(p_*)(-dp_*) =  (Q_1+Q_2)dt  = -{Q_1+Q_2\over \dot p_*}.
\end{equation}
Assuming power-law scalings so that the ratio of $Q_1$ to $Q_2$ is constant if follows that
\begin{equation}
S(p_*) \propto -{Q_2\over \dot p_*} \propto p_*^2 f(p_*) R^3.
\end{equation}
Further with these scalings
\begin{equation}
R^3 \propto p_*^{-6/(1+3\mu)}
\end{equation}
and because acceleration from the injection energy $p_0$ to $p_*$ is by definition fast so that there are no dilution effects,
\begin{equation}
f(p_*) =  f(p_0) \left(p_*\over p_0\right)^{-3U_1/(U_1-U_2)}.
\end{equation}
If we now use the two injection models discussed above we deduce that, in the case of a constant injection number fraction,
\begin{equation}
f(p_0) \propto p_0^{-3} \propto p_*^{-9/(1+3\mu)}
\end{equation}
and thus the effective high-energy source spectrum has exponent
\begin{equation}
{\partial \ln S(p_*)\over\partial \ln p_*}
= 2 -  {3U_1\over U_1 - U_2} + {3\over 1+3\mu}\left(U_1+2U_2\over U_1-U_2\right)
\end{equation}
which is excessively hard because of the very strong injection at early times.

More interesting is the case of self-regulated injection where the scaling 
\begin{equation}
f(p_0) \propto p_*^{6/(1+3\mu)}
\end{equation}
exactly cancels the $R^3$ term and we deduce
\begin{equation}
{\partial \ln S(p_*)\over\partial \ln p_*}
= 2 - {3U_1\over U_1 - U_2}.
\end{equation}
In this case the escaping spectrum is the same as the production spectrum and energy is equi-distributed across both.  A similar conclusion is reached by  \cite{Caprioli2010160}.

In reality of course these toy models are far too crude and as noted above they should only be taken as an indication of some of the effects that can occur and as a guide to further numerical studies.   They do however strongly suggest that the reason the spectrum softens between the ``knee'' region and the ``ankle'' (if this is a source effect and not a propagation effect) is that the acceleration responsible for these particles occurs early in the remnant evolution at a time when the shock is very fast, but not at full power.  Prior to sweep-up the bulk of the explosion energy is in the form of kinetic energy and is not available for acceleration.  Thus although it is possible to accelerate to very high energies, the total power is limited, and thus the source spectrum must turn down in this region.   The smooth matching onto the standard spectrum at lower energies then follows naturally from the shock and remnant dynamics.

\section{Acknowldegments}

The author wishes to thank les A\'eroport de Paris for providing power for his laptop and a comfortable working environment in the departure lounge while writing the first outline of this paper.  He is also grateful to the referee for an unusually thorough and insightful report which has substantially improved the presentation.
%
%
%
%

\bibliographystyle{apj}
\bibliography{mn-jour,escaperefs}

\begin{thebibliography}{26}
\expandafter\ifx\csname natexlab\endcsname\relax\def\natexlab#1{#1}\fi

\bibitem[{{Achterberg} {et~al.}(1994){Achterberg}, {Blandford}, \&
  {Reynolds}}]{1994A&A...281..220A}
{Achterberg}, A., {Blandford}, R.~D., \& {Reynolds}, S.~P. 1994, \aap, 281, 220

\bibitem[{{Axford} {et~al.}(1977){Axford}, {Leer}, \&
  {Skadron}}]{1977ICRC...11..132A}
{Axford}, W.~I., {Leer}, E., \& {Skadron}, G. 1977, in International Cosmic Ray
  Conference, Vol.~11, International Cosmic Ray Conference, 132--137

\bibitem[{{Bell}(1978)}]{1978MNRAS.182..147B}
{Bell}, A.~R. 1978, \mnras, 182, 147

\bibitem[{{Bell}(2004)}]{2004MNRAS.353..550B}
---. 2004, \mnras, 353, 550

\bibitem[{{Blandford} \& {Ostriker}(1978)}]{1978ApJ...221L..29B}
{Blandford}, R.~D., \& {Ostriker}, J.~P. 1978, \apjl, 221, L29

\bibitem[{Caprioli {et~al.}(2010)Caprioli, Amato, \& Blasi}]{Caprioli2010160}
Caprioli, D., Amato, E., \& Blasi, P. 2010, Astroparticle Physics, 33, 160

\bibitem[{{Caprioli} {et~al.}(2010){Caprioli}, {Amato}, \&
  {Blasi}}]{2010APh....33..307C}
{Caprioli}, D., {Amato}, E., \& {Blasi}, P. 2010, Astroparticle Physics, 33,
  307

\bibitem[{{Casanova} {et~al.}(2010){Casanova}, {Aharonian}, {Fukui}, {Gabici},
  {Jones}, {Kawamura}, {Onishi}, {Rowell}, {Sano}, {Torii}, \&
  {Yamamoto}}]{2010PASJ...62..769C}
{Casanova}, S., {et~al.} 2010, \pasj, 62, 769

\bibitem[{{Draine} \& {McKee}(1993)}]{1993ARA&A..31..373D}
{Draine}, B.~T., \& {McKee}, C.~F. 1993, \araa, 31, 373

\bibitem[{{Drury} {et~al.}(1999){Drury}, {Duffy}, {Eichler}, \&
  {Mastichiadis}}]{1999A&A...347..370D}
{Drury}, L.~O., {Duffy}, P., {Eichler}, D., \& {Mastichiadis}, A. 1999, \aap,
  347, 370

\bibitem[{{Drury} {et~al.}(1996){Drury}, {Duffy}, \&
  {Kirk}}]{1996A&A...309.1002O}
{Drury}, L.~O., {Duffy}, P., \& {Kirk}, J.~G. 1996, \aap, 309, 1002

\bibitem[{{Drury} \& {Keane}(1995)}]{1995NuPhS..39..165D}
{Drury}, L.~O., \& {Keane}, A.~J. 1995, Nuclear Physics B Proceedings
  Supplements, 39, 165

\bibitem[{{Drury} {et~al.}(2003){Drury}, {van der Swaluw}, \&
  {Carroll}}]{2003astro.ph..9820O}
{Drury}, L.~O., {van der Swaluw}, E., \& {Carroll}, O. 2003, ArXiv Astrophysics
  e-prints

\bibitem[{{Eichler}(1979)}]{1979ApJ...229..419E}
{Eichler}, D. 1979, \apj, 229, 419

\bibitem[{{Ellison} \& {Eichler}(1984)}]{1984ApJ...286..691E}
{Ellison}, D.~C., \& {Eichler}, D. 1984, \apj, 286, 691

\bibitem[{{Gabici} \& {Aharonian}(2007)}]{2007ApJ...665L.131G}
{Gabici}, S., \& {Aharonian}, F.~A. 2007, \apjl, 665, L131

\bibitem[{{Gabici} {et~al.}(2007){Gabici}, {Aharonian}, \&
  {Blasi}}]{2007Ap&SS.309..365G}
{Gabici}, S., {Aharonian}, F.~A., \& {Blasi}, P. 2007, \apss, 309, 365

\bibitem[{{Ginzburg} \& {Ptuskin}(1981)}]{1981ICRC....2..336G}
{Ginzburg}, V.~L., \& {Ptuskin}, V.~S. 1981, in International Cosmic Ray
  Conference, Vol.~2, International Cosmic Ray Conference, 336--339

\bibitem[{{Krymskii}(1977)}]{1977DoSSR.234Q1306K}
{Krymskii}, G.~F. 1977, Akademiia Nauk SSSR Doklady, 234, 1306

\bibitem[{{Malkov}(1997)}]{1997ApJ...485..638M}
{Malkov}, M.~A. 1997, \apj, 485, 638

\bibitem[{{Malkov} {et~al.}(2005){Malkov}, {Diamond}, \&
  {Sagdeev}}]{2005ApJ...624L..37M}
{Malkov}, M.~A., {Diamond}, P.~H., \& {Sagdeev}, R.~Z. 2005, \apjl, 624, L37

\bibitem[{{Ohira} {et~al.}(2010){Ohira}, {Murase}, \&
  {Yamazaki}}]{2010A&A...513A..17O}
{Ohira}, Y., {Murase}, K., \& {Yamazaki}, R. 2010, \aap, 513, A17+

\bibitem[{{Ptuskin} \& {Zirakashvili}(2005)}]{2005A&A...429..755P}
{Ptuskin}, V.~S., \& {Zirakashvili}, V.~N. 2005, \aap, 429, 755

\bibitem[{{Reville} {et~al.}(2009){Reville}, {Kirk}, \&
  {Duffy}}]{2009ApJ...694..951R}
{Reville}, B., {Kirk}, J.~G., \& {Duffy}, P. 2009, \apj, 694, 951

\bibitem[{{V{\"o}lk} {et~al.}(2003){V{\"o}lk}, {Berezhko}, \&
  {Ksenofontov}}]{2003A&A...409..563V}
{V{\"o}lk}, H.~J., {Berezhko}, E.~G., \& {Ksenofontov}, L.~T. 2003, \aap, 409,
  563

\bibitem[{{Yamazaki} {et~al.}(2004){Yamazaki}, {Yoshida}, {Terasawa}, {Bamba},
  \& {Koyama}}]{2004A&A...416..595Y}
{Yamazaki}, R., {Yoshida}, T., {Terasawa}, T., {Bamba}, A., \& {Koyama}, K.
  2004, \aap, 416, 595

\end{thebibliography}

\appendix

\section{Escape probabilities in random walks}

Here we give the formal proof that a uniformly diffusing particle in three dimensions, if released at distance $R_1$ from the origin, will escape to infinity without ever entering a sphere of radius $R_0$ centred on the origin with probability $1- R_0/R_1$ and will enter the sphere at least once with probability $R_0/R_1$.  We use a standard technique in random walk theory and solve the steady state diffusion equation for a source located at $R_1$  and with an absorbing boundary condition on the sphere of radius $R_0$.  The ratio of the flux being absorbed at $R_0$ to that escaping to infinity is then the ratio of the respective probabilities as long as the random walk is well represented by a diffusion process (this will be the case as long as the length scales are large compared to the scattering mean-free path $\lambda$,  that is $R_1 - R_0\gg \lambda$).

The steady-state diffusion equation in spherical coordinates is
\begin{equation}
{1\over r^2}{\partial\over\partial r}\left(r^2\kappa {\partial f\over\partial r}\right) = \delta(r-R_1)
\end{equation}
where as usual $\delta$ denotes the Dirac delta distribution.  We need to solve this with boundary conditions $f(R_0)=0$ and $f\to 0$ as $r\to\infty$.   Away from the source the steady-state diffusion equation is trivially integrable to give
\begin{equation}
f(r) = A \int{dr\over r^2\kappa} + B
\end{equation}
where $A$ and $B$ are constants of integration.   Considering the simple case where $\kappa$ is a constant this gives
\begin{equation}
f = B - {A\over\kappa} {1\over r}
\end{equation}
and thus the solution is, for  $R_0<r<R_1$,
\begin{equation}
f = C \left({1\over R_0} - {1\over r}\right)
\end{equation}
 where $C = A/\kappa$ is another constant and
\begin{equation}
f= {R_1\over r}  C \left({1\over R_0} - {1\over R_1}\right) = {C\over r} \left({R_1\over R_0} - 1\right)
\end{equation}
for $r>R_1$.

It follows immediately that the fluxes being absorbed at the inner boundary and escaping to infinity are in the ratio $1$ to $R_1/R_0 - 1$ and thus that the probability of absorption at the inner boundary is $R_0/R_1$ and of escape to infinity $1-R_0/R_1$.

Note that the calculation can easily be generalised to the case where $\kappa$ is non-constant and space has $n$ dimensions.  In this case the solution inside the injection radius is just
\begin{equation}
f(r) =  A \int_{R_0}^r {dr'\over r'^{n-1} \kappa(r')}
\end{equation}
and that outside is
\begin{equation}
f(r) = A \int_{R_0}^{R_1} {dr'\over r'^{n-i} \kappa(r')}
\int_r^\infty {dr'\over r'^{n-1} \kappa(r')} \left(\int_{R_1}^\infty {dr'\over r'^{n-i} \kappa(r')}\right)^{-1}
\end{equation}

It follows that the probability of return to $R_0$ if released at $R_1$ is 
\begin{equation}
\int_{R_1}^\infty {dr'\over r'^{n-i} \kappa(r')} \left(\int_{R_0}^\infty {dr'\over r'^{n-i} \kappa(r')}\right)^{-1}
\end{equation}
and of escape
\begin{equation}
\int_{R_0}^{R_1} {dr'\over r'^{n-i} \kappa(r')} \left(\int_{R_0}^\infty {dr'\over r'^{n-i} \kappa(r')}\right)^{-1}
\end{equation}

As is intuitively clear escape is easier the more spatial dimensions are available and the more $\kappa$ is an increasing function of radius, but it is interesting that this can be quantified in such simple closed formulae.

\end{document}